Perspective

**Title:** Epigenetic regulation of repetitive DNA through mitotic asynchrony following double fertilization in angiosperms

**Author:** John D. Laurie

**Affiliation:** Center for Plant Science Innovation, University of Nebraska, Lincoln, Nebraska 68588-0660, USA.

**Correspondence to:** jlaurie2@unl.edu or johndlaurie3@gmail.com

**Abstract**

Several recent studies show that companion cells in flowering plant gametophytes relax epigenetic control of transposable elements (TEs) to promote production of small RNA that presumably assist nearby reproductive cells in management of TEs. In light of this possibility, a closer look at the timing of cell division in relation to angiosperm double fertilization is warranted. From such an analysis, it is conceivable that double fertilization can drive angiosperm evolution by facilitating crosses between genetically diverse parents. A key feature of this ability is the order of cell division following double fertilization, since division of the primary endosperm nucleus prior to the zygote would produce small RNA capable of identifying TEs and defining chromatin states in the zygote prior to its entry into S-phase of the cell cycle. Consequently, crosses leading to increased ploidy or between genetically diverse parents would yield offspring better capable of managing a diverse complement of TEs and repetitive DNA. Considering double fertilization in this regard challenges previous notions that the primary purpose of endosperm is for improved seed reserve storage and utilization.

**Introduction**

The origin of angiosperms has been a topic of considerable interest for more than a century. Since their emergence, angiosperms have exploded in diversity and colonized much of the planet, far outnumbering all other plant species. Charles Darwin referred to the rapid emergence of angiosperms as an "abominable mystery" (Friedman, 2009) and the driving force behind this success remains a mystery. One approach to solving this mystery is to scrutinize features that are unique to angiosperms, such as endosperm. In angiosperms, double fertilization involves union of two sperm cells from a single pollen grain with the egg and the adjacent central cell in a reduced megagametophyte also known as the embryo sac. Double fertilization therefore gives rise to two entities; the embryo which goes on to form a plant, and an endosperm that plays a supportive role in seed development and germination (Berger et al., 2008).

Evolution of endosperm and its contribution to angiosperm success has been an enigma since the discovery of double fertilization more than 100 years ago. Two favored explanations for this evolution are the sexualization of the megagametophyte and the altruistic embryo hypotheses (Sargant, 1900; Strasburger, 1900; Coulter, 1911). With both of these models, endosperm is described as a tissue that evolved to replace the megagametophyte as primary support for the embryo. For species with large endosperms containing significant storage products a supportive role is obvious, but for many species from diverse taxa endosperm is minimal. Given the diversity of angiosperms as a whole and the fact that endosperm never develops beyond a few cells for many species, it is odd that endosperm has so rarely been lost. This implies that endosperm plays an essential role for angiosperms separate from seed nutrient storage. One developmental stage where endosperm has the potential to provide a significant advantage is immediately following fertilization.

Although a clear advantage for having evolved endosperm from megagametophyte has not been proven, aberrant endosperm development is often observed in interspecific and interploidy crosses (Josefsson et al., 2006; Ishikawa et al., 2011). This leads one to think that seed developmental programming is sensitive to parental genetic and/or epigenetic interactions (Gutierrez-Marcos et al., 2003; Kinoshita, 2007). Possible explanations to observed postzygotic barriers include parental genome imbalance and negative genome interactions leading to genome shock (Bushell et al., 2003; Dilkes and Comai, 2004). Our understanding of these events is becoming clearer due mainly to recent studies utilizing high resolution analyses of DNA methylation in endosperm and companion cells in both male and female gametophytes (Calarco et al., 2012; Ibarra et al., 2012). A common theme in these studies is the importance of epigenetic processes (Kohler et al., 2012). What is lacking, though, is a comprehensive model outlining how endosperm evolved and what special role it could play based on current knowledge of how RNA silencing influences epigenetics over the course of mitotic cell division. Here, I present this perspective and describe how endosperm could facilitate creation of genetic diversity by providing flowering plants with a robust means to cope with heterochromatin and diverse TEs during sexual reproduction. Emphasis is placed on very early events in seed development where a significant advantage can be gained through enhanced epigenetic regulation. Although speculative, the model addresses several features of endosperm from diverse angiosperm taxa and outlines how endosperm may have evolved in a gymnosperm ancestor.

**RNA silencing, epigenetics and the cell cycle**

Recent work on model organisms has shed light on how epigenetic mechanisms acting on chromatin function throughout the cell cycle (Chen et al., 2008; Kloc et al., 2008). Chromatin can be broadly classified as active euchromatin that is gene-rich and silent heterochromatin that is often gene-poor due largely to the abundance of TEs, TE remnants, and other repetitive DNA. The basic unit of chromatin is the nucleosome, where roughly 147 bp of DNA is wrapped around 8 core histone proteins; two each of H2A, H2B, H3 and H4. Covalent modifications to histone tails control interactions of histones with DNA and other proteins that collectively influence both higher order chromatin structure and access of the transcription and recombination machinery. Heterochromatin is characterized by repressive signatures that include specific histone modifications and DNA methylation on cytosine residues (Law and Jacobsen, 2010; Beisel and Paro, 2011). Additionally, heterochromatin is often sequestered in discrete nuclear domains and replicates separately from euchromatin late in S-phase of the mitotic cell cycle (Akhtar and Gasser, 2007; Lee et al., 2010). In budding yeast, histone tails are modified during S-phase to ensure that epigenetic programming is maintained in daughter cells (Rusche et al., 2003). Large-scale rewriting of epigenetic programming also occurs primarily during cell division and to a lesser extent in non-dividing cells. In fission yeast, silent heterochromatic DNA is paradoxically transcribed to ensure both the establishment and maintenance of its silent state. Interestingly, transcription of heterochromatin occurs during S-phase of the cell cycle and utilizes components of the RNA silencing pathway for its subsequent silencing (Chen et al., 2008; Kloc et al., 2008). This process is especially important at pericentromeric repetitive DNA where transcription coupled to RNA silencing directs heterochromatin formation by recruiting Clr4 to methylate histone H3 on lysine 9 (H3K9me) (Buhler and Moazed, 2007). H3K9me is subsequently recognized by HP1-like chromodomain proteins that package the DNA into repressive chromatin, establishing functional centromeres leading into mitosis (Motamedi et al., 2008). In establishing repressive or heterochromatic DNA at repetitive loci, cells prevent TE escape by suppressing unwarranted transcription and recombination that would ultimately compromise genome integrity. The timing and nature of heterochromatin formation appears to be tightly controlled in dividing cells, thus protecting euchromatin from TE escape that could occur during DNA replication.

RNA silencing is present in diverse organisms and has been studied intensively over the past decade (Fulci and Macino, 2007; Eamens et al., 2008; Batista and Marques, 2011; Ketting, 2011). In plants, the RNA silencing mechanism has diversified into several pathways utilizing specific and redundant components (Eamens et al., 2008). Simplistically, double stranded RNA (dsRNA) is cleaved by Dicer-like (DCL) endonucleases into 20-24 nucleotide (nt) RNA duplexes. One of the small interfering RNA strands (siRNA) is then incorporated into an Argonaute (AGO) protein that guides either interference of complementary mRNA through posttranscriptional gene silencing (PTGS) or guides components of repressive chromatin to loci expressing complementary sequences resulting in transcriptional gene silencing (TGS). DsRNA can form in cells through the activity of viruses or TEs, or by convergent transcription or transcription of RNA that folds to form secondary structures resembling hairpins. Additionally, RNA-dependent RNA polymerases (RDR) can act on RNA converting it into dsRNA. RNA silencing, therefore, plays an important role in modulating gene expression through PTGS and on maintaining genome integrity by controlling active genetic elements with TGS and PTGS. In genomes of organisms where large stretches of repetitive DNA occur, RNA silencing assists in controlling such DNA by cleaving RNA generated from the loci and by helping to recruit components of repressive chromatin. RNA silencing is, therefore, an important player in the functioning of chromosomes, especially at centromeres where repetitive DNA is abundant. When RNA silencing is compromised in diverse organisms, activation of TEs and difficulty in cell division are often observed due largely to failure in establishing heterochromatin near centromeres (Lippman et al., 2004).

In Arabidopsis DCL3 and AGO4 facilitate TGS by guiding DNA methylation at cytosine residues in a pathway referred to as RNA-directed DNA methylation (RdDM) (Law and Jacobsen, 2010). Such loci are also targets for KRYPTONITE that performs H3K9 methylation leading to chromatin that is for the most part transcriptionally silent (Johnson et al., 2002). The plant specific RNA polymerases Pol IV and Pol V transcribe loci that are both initiation sites and targets for 24-nt siRNA AGO4-directed TGS (Wierzbicki et al., 2008). Interestingly, RNA Pol V transcribes loci that can either act independently or together with RNA Pol IV-derived 24-nt siRNA in establishing repressive chromatin. Recent work suggests that RNA Pol V might be coupled to DNA replication and thus active during S-phase of the cell cycle, resembling TGS in fission yeast (Pontes et al., 2009). The Arabidopsis RdDM pathway is especially active at Gypsy-like retrotransposons that populate pericentromeric regions and at the edges of long TEs throughout the genome (Zemach et al., 2013). When RdDM is compromised in plants, a shift to PTGS can be seen (Tanurdzic et al., 2008; Slotkin et al., 2009). The hallmark of such a shift is a reduction in TGS-associated 24-nt siRNA that is accompanied by an increase in 21-nt siRNA active in PTGS. Hence, multiple layers of epigenetic regulation have evolved in plants to suppress transcription of repetitive DNA. Consequently, repetitive DNA is usually sequestered into repressive heterochromatin that is heavily methylated in the CpG context and supported by back-up RNA silencing machinery capable of both TGS and PTGS. For further protection, an important feature of RNA silencing in plants is its ability to spread from initiating cells to neighboring cells and systemically to distal parts of the plant. Recent studies show that all sizes of siRNA (21, 22 and 24 nt) are capable of moving and initiating both PTGS and TGS in distant cells (Dunoyer et al., 2010; Molnar et al., 2010). Such ability provides plants with a robust defense mechanism enabling systemic protection against mobile genetic elements in the form of TEs and viruses.

**The Angiosperm embryo sac at fertilization**

Double fertilization and allocation of seed reserves post-fertilization are features common to all flowering plants. Since both of these features are also present in the genus *Gnetum*, a small group of nonflowering seed plants, it is likely that they were present in the predecessor to angiosperms (Friedman and Carmichael, 1996, 1998). Contrary to what occurs in *Gnetum*, however, where two embryos result from double fertilization, angiosperms produce an embryo and an endosperm (Fig. 1). It, therefore, seems that special circumstances exist around the events of double fertilization in angiosperms that coordinate formation of an embryo and an endosperm and essentially force the endosperm to maintain ancient megagametophyte programming. An important component to this process is very likely the epigenetic state of the central cell. As demonstrated in maize, rice and Arabidopsis, demethylation of genomic DNA occurs prior to fertilization of the central cell (Gutierrez-Marcos et al., 2006; Gehring et al., 2009; Hsieh et al., 2009; Zemach et al., 2010b). In Arabidopsis, action of the 5-methylcytosine DNA glycosylase DEMETER causes the genome of the central cell to be reduced in CpG methylation at TEs, especially those residing near genes in euchromatin (Ibarra et al., 2012). The result is TE activation and aggressive RNA silencing that yields small RNAs corresponding to the demethylated TEs. A consequence of this process is differential methylation between maternal and paternal genomes following fertilization and the possible spread of TE-derived small RNA into the nearby egg or zygote (Kinoshita et al., 2004; Gehring et al., 2006; Jullien et al., 2006; Calarco et al., 2012; Ibarra et al., 2012). Methylation at CpG dinucleotides is normally abundant in repetitive DNA and heterochromatin (Hsieh et al., 2009; Zemach et al., 2013), and at genes, but with significant reduction at points of transcription start and termination (Zemach et al., 2010a). The DNA methyltransferase MET1 is primarily responsible for maintaining CpG methylation in Arabidopsis and plays an important role in suppressing TEs (Law and Jacobsen, 2010). In the central cell during megagametophyte development DEMETER activity is supported by suppression of the MET1 gene (Jullien et al., 2008; Jullien et al., 2012). The resulting global demethylation of the maternal genome occurs not only at repetitive loci, but also at a number of coding genes where demethylation of regulatory sequences results in parent-specific or imprinted expression in the endosperm (Gehring et al., 2009; Hsieh et al., 2009). In Arabidopsis, Polycomb-group proteins help enforce imprinting by directing histone H3 lysine-27 methylation of repressed alleles (Baroux et al., 2006). In addition to maternal and paternal chromatin differences, for most flowering plants two copies of the maternal genome participate in fertilization resulting in triploid endosperm. Thus, fertilization in the central cell involves two maternal genomes with reduced DNA methylation and apparently relaxed chromatin and one paternal genome with relatively compact chromatin. Studies also show that heterochromatin is relaxed in the vegetative nucleus of pollen and like in the central cell TEs become reduced in CpG methylation leading RNA silencing to produce TE-derived siRNA (Slotkin et al., 2009). These siRNA accumulate in sperm cells that also show reduced CpHpH methylation at pericentromeric regions (Calarco et al., 2012). Authors from these recent studies propose that companion cells (ie. central cell of megagametphyte and vegetative nucleus of pollen) relax epigenetic silencing of TEs to allow transcription coupled to RNA silencing to produce mobile small RNA that load into the nearby egg and sperm cells, thus providing offspring protection from possible negative consequences of TE activity (Martienssen, 2010; Mosher and Melnyk, 2010). It is conceivable that equal processes may similarly occur in most gymnosperms, since multicellular megagametophytes and tricellular pollen are also the norm. However, for these processes to have played a major role in angiosperm evolution there must be something special about these events that occurs only in angiosperms. By looking carefully at angiosperm reproduction two striking features have the potential for providing a profound influence on epigenetic processes. These are the bi-parental nature of endosperm and the order of cell division following double fertilization. Collectively, the conditions of the parental genomes in the primary endosperm nucleus are unique to angiosperms and set the stage for important events following fertilization.

**Mobile epigenetic signals provide an evolutionary advantage for double fertilization**

In light of the possibility that small RNA can be generated during cell division and are mobile, one must carefully consider the timing of cell division following double fertilization. When the endosperm proceeds to divide following fertilization it must construct functional centromeres, manage large heterochromatic domains and deal with relaxed epigenetic silencing at repetitive DNA that occurred in the central cell and sperm. Since RNA silencing plays such a pivotal role in proper centromere formation and establishment of heterochromatin, RNA silencing pathways are likely busy in the endosperm as it proceeds in cell division. Such processes are especially important when endosperm results from a distant cross. As populations of a species become separated by distance, reproductive isolation leads not only to allelic variation but also to differences in TE load and diversity as a consequence of the arms race or coevolution of TEs and host genome defenses (Aravin et al., 2007; Fedoroff, 2012). During S-phase of the cell cycle DNA at repetitive loci relaxes for replication which in turn produces a hemi-methylated DNA copy. To maintain genome stability epigenetic silencing must act quickly at these loci to prevent TE escape. DNA replication is therefore, risky business at loci housing TEs and possibly a contributing factor to the temporal segregation of euchromatin and heterochromatin DNA replication (Lee et al., 2010). Some of the challenges imposed by TEs include specification of heterochromatin and euchromatin and identification of the boundaries between them, and identifying centromeres and telomeres so that mitosis can proceed successfully. Fortunately, since the endosperm is short lived and does not pass its DNA on to the next generation, mutations caused by TE mobility during the cell cycle would not pose a significant threat to the embryo. As the primary endosperm nucleus proceeds with its first cell cycle, genome compatibility would be assessed by the ability of the epigenetic machinery to manage TEs and repetitive DNA. Small RNA involved in this process would assist the resulting daughter cells in maintaining genome integrity. Such information would likely emanate from all heterochromatic and repetitive loci including pericentromeric loci. Since the maternal genome likely controls early events during initial endosperm cell division , any diverse TEs introduced through the paternal genome that manage to escape transcriptional repression would subsequently be acted upon by the maternal PTGS pathway. As the endosperm continues to divide, small RNA signals would be amplified. Such a process would allow successful mitosis and is very likely how endosperm assesses and facilitates genome compatibility. In a similar manner, the genomic defenses of the zygote, which are likely mostly maternally-derived at fertilization, must also deal with paternal TEs during cell division. Unlike for endosperm though, the zygote cannot afford deleterious mutations and must aggressively prevent TE activation. Diverse TEs introduced through pollination by a distant relative, therefore, pose a serious risk to the zygote if unrecognized and not controlled during cell division.

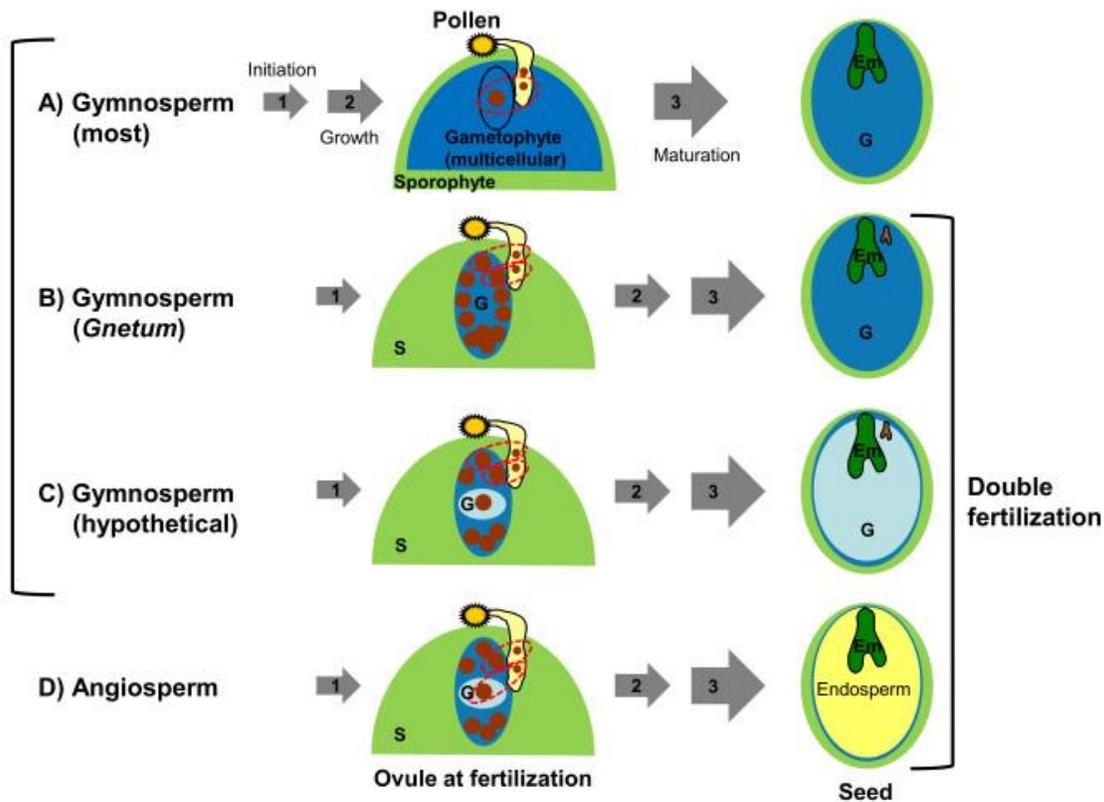

Fig. 1 Sexual reproduction in seed plants. A) In gymnosperms, an egg cell is fertilized by a single sperm cell from pollen in an archegonium supported by a large multicellular megagametophyte. Prior to fertilization initial gametogenesis (1) is followed by significant gametophyte growth and nutrient allocation (2). After fertilization both the embryo and gametophyte continue to grow and develop to produce a mature seed (3). B) In *Gnetum*, fertilization occurs in a megagametophyte reduced to a single cell containing several nuclei. Double fertilization takes place yielding twin embryos that initially develop together. As the seed matures one embryo aborts leaving a healthy embryo and a large supportive megagametophyte. C) In a hypothetical angiosperm predecessor, a cell of the megagametophyte undergoes extensive DNA demethylation (cell with light blue cytoplasm). Double fertilization occurs as in *Gnetum*, but all post-fertilization resource allocation occurs in the specialized cell of the megagametophyte that has undergone DNA demethylation. At seed maturity, only one embryo persists that is supported by the "specialized" megagametophyte. D) In angiosperm, the specialized cell of the megagametophyte that has undergone DNA demethylation gets fertilized along with the egg cell resulting in endosperm and an embryo. G, megagametophyte; S, sporophyte; Em, embryo.

    As the primary endosperm nucleus enters the cell cycle and replicates for the first time, S-phase epigenetic signals would be produced. It is at this point, where transcription, perhaps by

Pol II or IV, produces RNA that is acted upon by DCL and AGO proteins capable of directing both PTGS and TGS (Mosher et al., 2009; Mosher, 2010). The siRNA of all size classes emanating from repetitive loci throughout the genome would then, through mobility, be capable of performing their respective PTGS and TGS duties in adjacent or distal cells. Of interest are the recently identified loci showing reduced CpG methylation at euchromatic TEs in the central cell and reduced CpHpH methylation at pericentromeric repeats in the sperm (Calarco et al., 2012; Ibarra et al., 2012). Such loci would have a greater potential to be transcribed and yield small RNA. Additionally, in Arabidopsis DNA methyltransferases are weakly, if at all, expressed in endosperm ensuring that small RNA can be continuously produced as endosperm nuclei divide (Jullien et al., 2012). For the embryo to benefit from these epigenetic signals, the endosperm would have to divide prior to the zygote entering or proceeding with the cell cycle. In this way, information learned about types and locations of TEs or other repetitive DNA, especially at centromeres and ones associated with genes in euchromatin, would be available at mitosis or at the initiation of chromatin relaxation during DNA replication in the zygote, thus helping with centromere formation and preventing mutagenesis resulting from TE escape. Such information might help to define the heterochromatin-euchromatin boundaries and provide tighter transcriptional or post-transcriptional control of TEs, possibly allowing sooner activation of the parental genomes. In support of this prediction, the primary endosperm nucleus appears to divide prior to the zygote in all flowering plants. In fact, nuclei in the endosperm often divide several times before the first division of the zygote, especially in some basal angiosperms (Table 1). Conceivably, the zygote fills with small RNA or other unidentified mobile epigenetic signals derived from cell division in the endosperm capable of directing chromatin modifications and combating transcriptionally active TEs (Fig. 2). For identical or very similar parents this may not be all that important, since the egg would already possess all or most of the epigenetic information specifying repetitive DNA, but for diverse parents such a process would prime the zygote by arming it with epigenetic information leading into its first S-phase of the cell cycle. Without double fertilization diverse TEs would therefore pose a problem for the zygote and likely be able to escape the maternal-specified genome defenses. In Drosophila, species-specific heterochromatin prevents mitotic chromosome segregation leading to hybrid lethality (Ferree and Barbash, 2009).

**Table 1**

**Number of mitotic divisions in endosperm prior to zygote mitosis**

| Group | Family | Genus | Divisions[a] | Reference |
|---|---|---|---|---|
| Eudicot | Brassicaceae | Arabidopsis | 5 | c |
| Magnoliid | Annonaceae | Annona | 2 | d |
|  | Piperaceae | Peperomia | At least 3 | h |
| Monocot | Poaceae | Zea | 3 | e |
| Basal | Amborellaceae | Amborella | Multiple[b] | f |
|  | Nymphaeaceae | Nuphar | multiple | f |
|  | Illiciaceae | Illicium | multiple | f |
|  | Hydatellaceae | Trithuria | 2 | g |

[a]Approximate number of endosperm nuclear division at time of zygote mitosis.
[b]Multicellular endosperm at time of first zygote division.
[c](Boisnard-Lorig et al., 2001)
[d](Lora et al., 2010)
[e](Mol et al., 1994)
[f](Floyd and Friedman, 2001)
[g](Rudall et al., 2009)
[h](Madrid and Friedman, 2010)

Double fertilization helps flowering plants deal with chromatin challenges by turning the endosperm into a cell cycle experiment where successful DNA replication and cell division in the presence of active TEs is monitored and facilitated by the RNA silencing machinery. Information about TEs is then passed from dividing endosperm nuclei to the zygote. Since endosperm is terminal in nature, detrimental effects caused by TE activity are not transmitted to offspring. This process allows for diverse genomes (from within a species and between closely related species) to succeed in reproduction yielding progeny with allelic diversity that otherwise would have failed due to TE/heterochromatin-related issues.

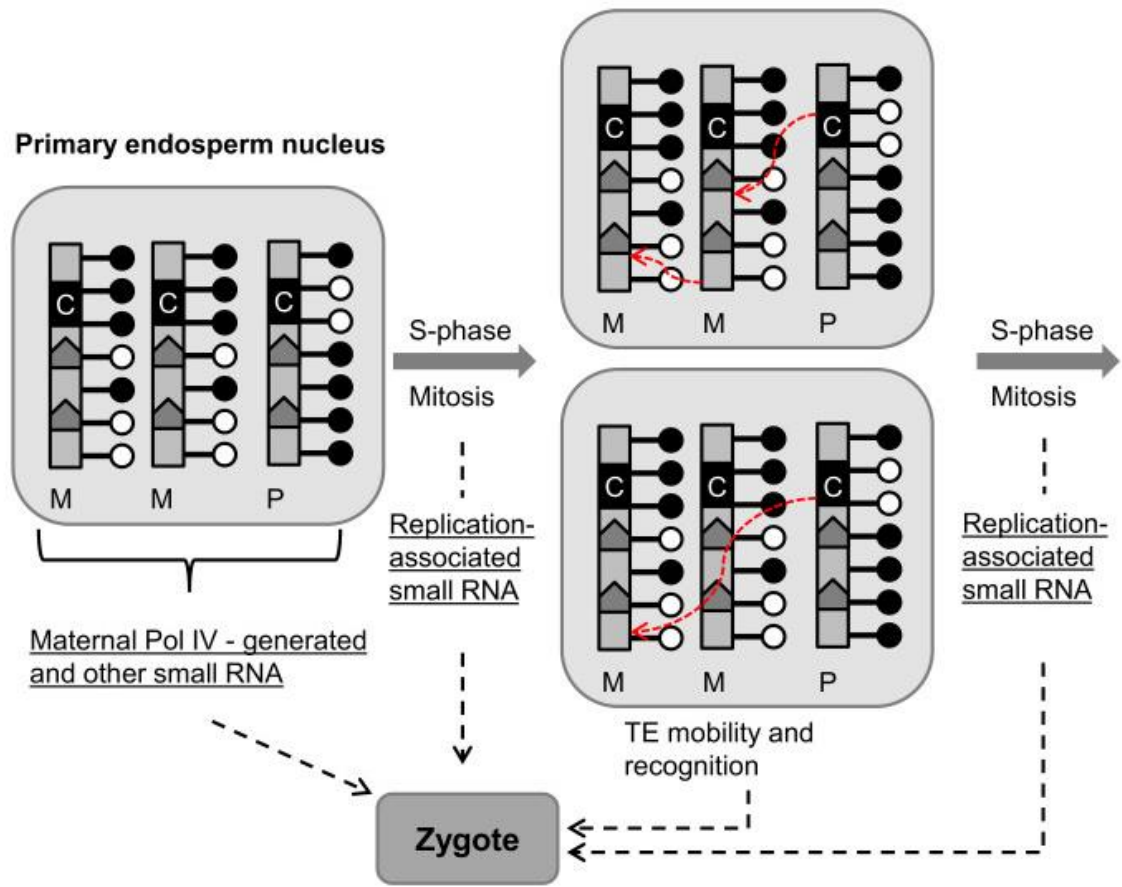

Fig. 2 Events associated with mitosis of the primary endosperm nucleus. The primary endosperm nucleus is a fertilization product derived from two copies of the maternal (M) genome and one copy of the paternal (P) genome that have been reduced in DNA methylation (open circles) at specific TEs prior to fertilization. Active paternal-genome derived TEs (red arrows) would preferentially land in the DNA methylation-reduced maternal genome. Small RNA produced from transcription of the TEs in dividing endosperm nuclei can conceivably relocate to specify both maternal and paternal chromatin states in the zygote (black arrows). C, centromere.

**Double fertilization in the hypothetical predecessor to angiosperms**

Angiosperms have a special form of double fertilization where reduced DNA methylation occurs in the central cell and possibly sperm prior to fertilization (Fig. 1). Initially, these could have occurred in gymnosperms in specialized companion or nurse cells of sporophytic or gametophytic origin as a means to reveal TEs to nearby gametes or the zygote (Fig. 3). In the megagametophyte of the predecessor to angiosperms, however, DNA demethylation would have occurred in a cell near the egg. In pollen of Arabidopsis, reduced CpG DNA methylation in the vegetative nucleus leads to TE activation and production of corresponding 21-nt siRNA that load into sperm cells (Slotkin et al., 2009). In the megagametophyte prior to angiosperm evolution, DNA demethylation likely became linked to the expression of megagametophyte developmental genes in a single megagametophyte cell, perhaps by simply up-regulating the DEMETER DNA glycosylase. Since all angiosperm embryo sacs are reduced to only a few cells, it is highly probable that seed storage responsibilities were assigned to this special, DNA methylation-reduced cell prior to angiosperm evolution. This step may have resolved conflicts with megagametophyte cells that had not undergone DNA demethylation. Coupling post-fertilization nutrient allocation to the demethylated companion cell and permitting it to participate in fertilization would approach what occurs in the central cell of angiosperms (Fig. 1). This is not hard to imagine since *Gnetum* species possess highly reduced megagametophytes that allocate seed reserves post-fertilization (Friedman and Carmichael, 1998). By having the responsibility of revealing TEs through genome-wide DNA demethylation together with post-fertilization allocation of seed storage reserves the ancestral companion cell would have been paramount for species survival. Fertilization by the second sperm cell of pollen capable of initiating embryo formation, as in the Gnetales, would have created a battle between the companion cell programming and that of the sperm cell. The initial battle over cell fate may constitute part of the imprinting phenomenon observed in endosperm today. Additionally, as a means to overcome this conflict of parental genomes many angiosperms have diverted seed nutrient reserve allocation away from the endosperm and assigned it to other cell types, including the cotyledons of the embryo for some species (like in the Fabaceae) and the perisperm of others (as in the Piperaceae and Hydatellaceae).

Reduced DNA methylation in the central cell would have created a relaxed genome that might have allow transcription machinery, either RNA Pol II or Pol IV, greater access to genes or loci that would normally have been suppressed over much of the cell cycle. Furthermore, ease of transcription might provide the female with tighter control over events following fertilization. This could also be a reason for including two or more copies of the female genome in endosperm. Two copies of relaxed chromatin could potentially express factors that would titrate out opposing factors introduced by the participating sperm cell (Lu et al., 2012). This might be important for processes involved in cell division. Interploidy crossed that upset the balance of maternal to paternal genome contribution have a significant impact on cell cycle related processes. Evidence from Arabidopsis and maize shows that when the maternal contribution to endosperm is increased precocious endosperm development is observed, whereas when the paternal contribution is increased endosperm development is delayed or prevented following a period of extended mitotic cell divisions (Leblanc et al., 2002; Li and Dickinson, 2010; Lu et al., 2012). This implies that the maternal genome contributes factors that promote development, while the paternal genome expresses factors that act to force cell division to proceed. Collectively this makes sense if the ancestral companion cell was indeed a megagametophyte cell containing relaxed chromatin with demethylated DNA programmed to activate TEs and pass epigenetic information on to the egg cell. It would be important to make sure that such a cell did not develop beyond its intended role and divert resources from the egg. Strict developmental programming would have been in place to ensure that the cell could not develop into an embryo, since TE activation would have caused detrimental mutations.

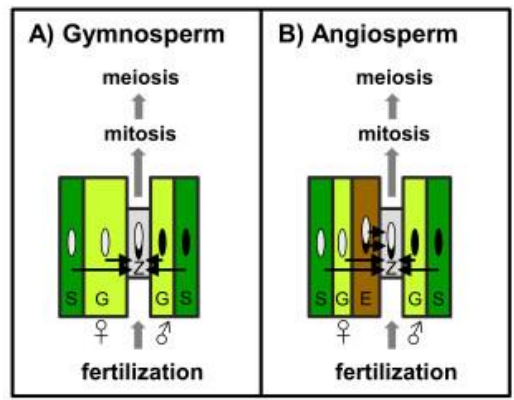

**Fig. 3** Potential routes for parental epigenetic signals to influence the progeny of seed plants. A) A simplistic representation of the reproductive cycle in gymnosperms where epigenetic silencing in the zygote (Z) is supported by adjacent gametophytic (G) and sporophytic (S) cells. B) In angiosperms, endosperm (E) may act as an additional source of epigenetic information capable of passing information from both parents to the zygote.

**A role for the maternal genome in identifying paternal TEs**

A tantalizing possiblility is that the demethylated maternal genome facilitates recognition of paternally-derived TEs during cell division (Fig. 2). With the maternal genome having relaxed chromatin and the pericentromeric repeats of the paternal genome being highly reduced in CpHpH methylation, any paternal TE that happens to escape maternal silencing defenses during cell division could potentially land in the maternal genome. In fact, with reduced maternal DNA methylation, active paternal TEs during cell division might preferentially land in the relaxed maternal chromatin (Dietrich et al., 2002; Pan et al., 2005; Liu et al., 2009). In Arabidopsis, RNA Pol IV associated 24-nt siRNA are generated solely from the maternal endosperm genome (Mosher et al., 2009). This suggests that the maternal genome is licensed for transcription by RNA Pol IV and that any newly introduced TEs could be processed and directed into the pool of small RNA capable of travelling to the zygote. Support for this notion comes from a study where DNA methylation was measured and showed a reduction of CpG methylation in the maternal endosperm genome that correlated with an increase in CpHpH methylation in the developing embryo (Hsieh et al., 2011). The authors of the study suggested that loss of CpG methylation permitted transcription that lead to an increase in corresponding 24-nt siRNA that were presumed to direct TGS in the embryo. This notion has gained further support recently when a central cell expressed miRNA was observed to silence an egg specific transgene (Ibarra et al., 2012). So in a cell with maternally-derived genome defenses, difficulties in controlling diverse TEs introduced through the paternal genome is overcome since any TE landing in the maternal genome would be recognized during cell division, and fed into the stream of mobile epigenetic information. Having two copies of the maternal genome would be beneficial to such a process since active paternal TEs would be more likely to land in the maternal genome. Additionally, multiple maternal genomes would reduce the likelihood of paternal TEs disrupting genes essential for endosperm development. Extreme versions of this process may be represented in the Piperaceae and Plumbaginaceae where eight and four copies of the maternal genome, respectively, participate in fertilization in the central cell. Essentially, the maternal genome of endosperm may act like the flamenco locus in flies, where active TEs that land in the flamenco locus are transcribed and converted into mobile RNA silencing signals that travel to nearby germ cells to enforce silencing (Brennecke et al., 2007).

For genomes that contain a diverse complement of TEs and repetitive DNA, placement of silent chromatin marks must be coordinated in an efficient manner to prevent TE escape. CpG methylation is used to silence the majority of repetitive DNA in plants. Defects causing depletion of CpG methylation at TEs leads to production of corresponding siRNA that then work through the RdDM defense pathway. DNA demethylation occurring in the central cell may produce mobile siRNA that reinforce or reset the boundaries between repeats and genes in euchromatin and at the boundaries of repetitive DNA throughout the genome of the zygote. Such a process could ensure that diverse TEs were controlled and not allowed to perturb nearby genes. Interestingly, when DNA methylation pathway mutants are restored to normal, a stepwise resetting of DNA methylation occurs at each pass through sexual reproduction (Teixeira et al., 2009). Sexual reproduction, therefore, plays a special role in epigenetic reprogramming. What role endosperm plays in this process is unknown. Additionally, how this process differs between angiosperms and gymnosperms remains to be seen. Preliminary studies comparing gymnosperms to angiosperms suggest that differences in small RNAs are apparent, but that types of DNA methylation and histone modifications may be conserved (Muller et al., 2012). It is tempting to think that angiosperms have an enhanced ability to recognize repetitive DNA and therefore are better able at defining euchromatin and heterochromatin boundaries that would allow greater success for distant crosses and in polyploidy genomes. A role in this process for endosperm and RNA polymerases seems logical and when additional information about RNA silencing and chromatin biology from gymnosperms is obtained important differences will be revealed (Dolgosheina et al., 2008; Morin et al., 2008).

**Implications of the mitotic asynchrony model**

The model presented here has several implications for the early steps in angiosperm evolution and suggests that certain criteria needed to be in place prior to endosperm evolution. For the process to have evolved, the predecessor to angiosperms would have possessed RNA silencing-directed TE control that produced mobile small RNA capable of inducing TGS and PTGS in distant cells. With this ability, a megagametophytic cell near the egg cell would be capable of revealing maternal TEs by active genome-wide DNA demethylation, perhaps by a DNA glycosylase similar to DEMETER working in concert with repression of MET1. Cell division or endoreduplication of such a cell would enable RNA Pol II, IV or V to transcribe RNA templates for production of siRNA capable of directing TGS and PTGS in the egg/zygote. Division of labour between RNA Pol IV and V may have relevance to the epigenetic signals surrounding fertilization and could assist the zygote with receipt or processing of epigenetic information before it proceeds in cell division. What rudimentary components of this process exist in extant non-flowering seed plants remains to be determined. A logical place to look for such events would be the enlarged cells of the archegonia of basal seed plants and cells of the chalazal megagametophyte of the Gnetales that undergo nuclear fusion prior to megagametophyte maturation (Friedman and Carmichael, 1998). Another implication of this model is the control over timing of the cell division. The model predicts that the endosperm may possess some control over when the zygote proceeds in cell division. Perhaps a component of the mobile epigenetic signaling emanating from the endosperm removes a factor promoting cell cycle delay in the zygote. This would ensure that cell division proceeds first in the endosperm and that the zygote has received protective signals prior to cell division. A requirement for the process would be that following fertilization the companion cell (endosperm nucleus) would have to proceed into S-phase of the cell cycle. Minimally, the cell would have to divide only once to promote DNA replication-associated transcription and possibly TE escape. Alternatively, the cell could achieve nearly the same effect through endoreduplication. In this case, the fertilized cell could simply undergo several rounds of endoreduplication to facilitate replication-associated transcription of heterochromatic DNA. However, endoreduplication may miss an important step in centromere recognition that occurs during cell division. With this in mind, reexamination of endosperm in diverse angiosperms is warranted. Angiosperms with no apparent endosperm development, as in the orchids for example, need to be reevaluated for signs of cell division or endoreduplication of the primary endosperm nucleus following double fertilization. Additionally, double fertilization may have evolved into a two-step process for some plants. It has been noted that for some angiosperms, including Arabidopsis, nuclei of the sperm and egg enter fertilization in G2 of the cell cycle (Friedman, 1999; Tian et al., 2005). In essence, the zygote is a polyploidy and the first step in the cell cycle for these plants is mitosis. For such plants, the first cell division may simply involve chromatin condensation and centromere recognition. Dealing with TEs in the zygote is therefore postponed until the second cell division when DNA is relaxed for replication. Prolonging the first S-phase of the zygote may ensure that the endosperm has divided several times and may also recruit the newly formed suspensor to assist the endosperm in epigenetic signaling to the zygote. Allowing gametes to enter fertilization in G2 would not only allow the suspensor to assist endosperm, but would put the suspensor in a position to adopt many of the roles endosperm might play in epigenetic signaling. A closer look at fertilization and suspensors in the orchid family is warranted to determine if this evolutionary step has indeed taken place. Lastly, this final S-phase that occurs just prior to fertilization in Arabidopsis has implication for the state of DNA methylation in the sperm cells. It remains to be determined if the reduced CpHpH methylation at pericentromeric repeats is restored before fertilization or if the repeats enter fertilization hypomethylated (Calarco et al., 2012).

Considering endosperm in this regard provides an alternative version of the sexualization of the megagametophyte model for endosperm evolution. Instead of being a tissue for improved nutritional support for the developing embryo a strong case can be made for endosperm evolving as a tissue to facilitate compatibility between diverse genomes. The model places importance on the initial cell divisions of endosperm, where a significant epigenetic advantage can be gained. Subsequent endosperm development should then be considered to follow more-or-less the ancient megagametophte programming, but with the addition of paternal input. As alluded to previously, this subsequent endosperm development is vulnerable to parental contribution (Leblanc et al., 2002; Josefsson et al., 2006; Li and Dickinson, 2010). Interspecific crosses and crosses between parents of different ploidy are known to disrupt endosperm development and often lead to reduced seed viability or aborted seeds. It is probable that during the early steps in evolution of endosperm, fail-safe mechanisms were put in place to prevent extremely diverged genomes from partaking in fertilization and wasting resources on potentially unfit progeny. In interspecific or interploidy crosses, that normally fail, artificially increasing the maternal contribution can increase endosperm success and production of viable seeds (Bushell et al., 2003). Early angiosperms may have used this phenomenon to their advantage. Fusion of several DNA methylation-reduced megagametophyte nuclei prior to fertilization would ensure that endosperm development could proceed in the presence of active TEs and foreign paternal heterochromatin. Early in angiosperm evolution the number of nuclei participating in central cell formation could have been reduced to an optimal number to counter the enhanced outcrossing ability. A braking system was therefore put in place in this two-step evolution process. For angiosperms where multiple nuclei fuse to form the central cell prior to fertilization it remains to be seen what advantage this feature might have for interspecific and interploidy crosses in the wild. Since most angiosperms are reported to allow only two nuclei to make up the central cell and participate in fertilization, it would appear that fewer may be more advantageous. Natural selection ensured that endosperm provided the necessary support in dealing with TEs, but that it did not allow uncontrolled crossing that would have jeopardized the species as a whole. Thus endosperm became the deciding factor for interspecific crosses.

**Summary**

Unique cell-cycle generated small RNA produced by dividing endosperm nuclei can conceivably direct epigenetic changes in the zygote of flowering plants. In this model, relaxed TE silencing through reduced DNA methylation facilitates transcription and production of corresponding small RNA. Additionally, the methylation-reduced maternal genome has the potential to attract active paternal-genome derived TEs and through the activity of RNA Pol IV, and possibly Pol II, produce mobile epigenetic signals specifying repetitive DNA from both parents. This information is most beneficial when the zygote divides after receiving the epigenetic information. This mitotic asynchrony model helps address several aspects of seed reproductive biology, including why the endosperm divides prior to the zygote, why the central cell of the megagametophyte and sperm selectively demethylate DNA, why a benefit can be obtained from allowing multiple maternal nuclei to partake in fertilization producing endosperm, and why endosperm can be reduced to only a few cells in certain species yet not so easily eliminated. Furthermore, an epigenetic advantage can be seen for entering fertilization in G2 of the cell cycle. With large-scale sequencing projects well underway, increased knowledge of TEs and genome defenses in diverse taxa will help to clarify the key evolutionary steps taken by angiosperms in dealing with repetitive DNA. The model provides a starting point for looking into genomic defenses and how they work over the cell cycle in and around reproductive cells of all seed plants. Additionally, the model forces us to rethink the initial purpose of endosperm and to consider how coevolution of TEs together with this added level of genome defense has led to the diversity of flowering plant and their seeds. Future work in this regard will help us gain a better understanding of angiosperm diversity and facilitate seed improvements for agricultural purposes.

**Acknowledgements**

The author thanks Rebecca Mosher (University of Arizona) and Frederic Berger (Temasek Life Sciences Laboratory) for discussions and helpful suggestions on improving the manuscript. The author is grateful to Brian A. Larkins (University of Nebraska) for his guidance and support.

**References**

**Akhtar, A., and Gasser, S.M.** (2007). The nuclear envelope and transcriptional control. Nature reviews. Genetics **8,** 507-517.
**Aravin, A.A., Hannon, G.J., and Brennecke, J.** (2007). The Piwi-piRNA pathway provides an adaptive defense in the transposon arms race. Science **318,** 761-764.
**Baroux, C., Gagliardini, V., Page, D.R., and Grossniklaus, U.** (2006). Dynamic regulatory interactions of Polycomb group genes: MEDEA autoregulation is required for imprinted gene expression in Arabidopsis. Genes & development **20,** 1081-1086.
**Batista, T.M., and Marques, J.T.** (2011). RNAi pathways in parasitic protists and worms. Journal of proteomics **74,** 1504-1514.
**Beisel, C., and Paro, R.** (2011). Silencing chromatin: comparing modes and mechanisms. Nature reviews. Genetics **12,** 123-135.
**Berger, F., Hamamura, Y., Ingouff, M., and Higashiyama, T.** (2008). Double fertilization - caught in the act. Trends in plant science **13,** 437-443.
**Boisnard-Lorig, C., Colon-Carmona, A., Bauch, M., Hodge, S., Doerner, P., Bancharel, E., Dumas, C., Haseloff, J., and Berger, F.** (2001). Dynamic analyses of the expression of the HISTONE::YFP fusion protein in arabidopsis show that syncytial endosperm is divided in mitotic domains. The Plant cell **13,** 495-509.
**Brennecke, J., Aravin, A.A., Stark, A., Dus, M., Kellis, M., Sachidanandam, R., and Hannon, G.J.** (2007). Discrete small RNA-generating loci as master regulators of transposon activity in Drosophila. Cell **128,** 1089-1103.

**Buhler, M., and Moazed, D.** (2007). Transcription and RNAi in heterochromatic gene silencing. Nature structural & molecular biology **14,** 1041-1048.

**Bushell, C., Spielman, M., and Scott, R.J.** (2003). The basis of natural and artificial postzygotic hybridization barriers in Arabidopsis species. The Plant cell **15,** 1430-1442.

**Calarco, J.P., Borges, F., Donoghue, M.T., Van Ex, F., Jullien, P.E., Lopes, T., Gardner, R., Berger, F., Feijo, J.A., Becker, J.D., and Martienssen, R.A.** (2012). Reprogramming of DNA methylation in pollen guides epigenetic inheritance via small RNA. Cell **151,** 194-205.

**Chen, E.S., Zhang, K., Nicolas, E., Cam, H.P., Zofall, M., and Grewal, S.I.** (2008). Cell cycle control of centromeric repeat transcription and heterochromatin assembly. Nature **451,** 734-737.

**Coulter, J.M.** (1911). The endosperm of angiosperms - Contributions from the Hull Botanical Laboratory 150. Bot Gaz **51,** 380-385.

**Dietrich, C.R., Cui, F., Packila, M.L., Li, J., Ashlock, D.A., Nikolau, B.J., and Schnable, P.S.** (2002). Maize Mu transposons are targeted to the 5' untranslated region of the gl8 gene and sequences flanking Mu target-site duplications exhibit nonrandom nucleotide composition throughout the genome. Genetics **160,** 697-716.

**Dilkes, B.P., and Comai, L.** (2004). A differential dosage hypothesis for parental effects in seed development. The Plant cell **16,** 3174-3180.

**Dolgosheina, E.V., Morin, R.D., Aksay, G., Sahinalp, S.C., Magrini, V., Mardis, E.R., Mattsson, J., and Unrau, P.J.** (2008). Conifers have a unique small RNA silencing signature. Rna-a Publication of the Rna Society **14,** 1508-1515.

**Dunoyer, P., Schott, G., Himber, C., Meyer, D., Takeda, A., Carrington, J.C., and Voinnet, O.** (2010). Small RNA duplexes function as mobile silencing signals between plant cells. Science **328,** 912-916.

**Eamens, A., Wang, M.B., Smith, N.A., and Waterhouse, P.M.** (2008). RNA silencing in plants: yesterday, today, and tomorrow. Plant physiology **147,** 456-468.

**Fedoroff, N.V.** (2012). PRESIDENTIAL ADDRESS Transposable Elements, Epigenetics, and Genome Evolution. Science **338,** 758-767.

**Ferree, P.M., and Barbash, D.A.** (2009). Species-specific heterochromatin prevents mitotic chromosome segregation to cause hybrid lethality in Drosophila. PLoS biology **7,** e1000234.

**Floyd, S.K., and Friedman, W.E.** (2001). Developmental evolution of endosperm in basal angiosperms: evidence from Amborella (Amborellaceae), Nuphar (Nymphaeaceae), and Illicium (Illiciaceae). Plant Syst Evol **228,** 153-169.

**Friedman, W.E.** (1999). Expression of the cell cycle in sperm of Arabidopsis: implications for understanding patterns of gametogenesis and fertilization in plants and other eukaryotes. Development **126,** 1065-1075.

**Friedman, W.E.** (2009). The meaning of Darwin's 'abominable mystery'. American journal of botany **96,** 5-21.

**Friedman, W.E., and Carmichael, J.S.** (1996). Double fertilization in gnetales: Implications for understanding reproductive diversification among seed plants. Int J Plant Sci **157,** S77-S94.

**Friedman, W.E., and Carmichael, J.S.** (1998). Heterochrony and developmental innovation: Evolution of female gametophyte ontogeny in Gnetum, a highly apomorphic seed plant. Evolution; international journal of organic evolution **52,** 1016-1030.

**Fulci, V., and Macino, G.** (2007). Quelling: post-transcriptional gene silencing guided by small RNAs in Neurospora crassa. Current opinion in microbiology **10,** 199-203.

**Gehring, M., Bubb, K.L., and Henikoff, S.** (2009). Extensive demethylation of repetitive elements during seed development underlies gene imprinting. Science **324,** 1447-1451.

**Gehring, M., Huh, J.H., Hsieh, T.F., Penterman, J., Choi, Y., Harada, J.J., Goldberg, R.B., and Fischer, R.L.** (2006). DEMETER DNA glycosylase establishes MEDEA polycomb gene self-imprinting by allele-specific demethylation. Cell **124,** 495-506.

**Gutierrez-Marcos, J.F., Pennington, P.D., Costa, L.M., and Dickinson, H.G.** (2003). Imprinting in the endosperm: a possible role in preventing wide hybridization. Philos T Roy Soc B **358,** 1105-1111.

**Gutierrez-Marcos, J.F., Costa, L.M., Dal Pra, M., Scholten, S., Kranz, E., Perez, P., and Dickinson, H.G.** (2006). Epigenetic asymmetry of imprinted genes in plant gametes. Nature genetics **38,** 876-878.

**Hsieh, T.F., Ibarra, C.A., Silva, P., Zemach, A., Eshed-Williams, L., Fischer, R.L., and Zilberman, D.** (2009). Genome-wide demethylation of Arabidopsis endosperm. Science **324,** 1451-1454.

**Hsieh, T.F., Shin, J., Uzawa, R., Silva, P., Cohen, S., Bauer, M.J., Hashimoto, M., Kirkbride, R.C., Harada, J.J., Zilberman, D., and Fischer, R.L.** (2011). Regulation of imprinted gene expression in Arabidopsis endosperm. Proceedings of the National Academy of Sciences of the United States of America **108,** 1755-1762.

**Ibarra, C.A., Feng, X., Schoft, V.K., Hsieh, T.F., Uzawa, R., Rodrigues, J.A., Zemach, A., Chumak, N., Machlicova, A., Nishimura, T., Rojas, D., Fischer, R.L., Tamaru, H., and Zilberman, D.** (2012). Active DNA demethylation in plant companion cells reinforces transposon methylation in gametes. Science **337,** 1360-1364.

**Ishikawa, R., Ohnishi, T., Kinoshita, Y., Eiguchi, M., Kurata, N., and Kinoshita, T.** (2011). Rice interspecies hybrids show precocious or delayed developmental transitions in the endosperm without change to the rate of syncytial nuclear division. The Plant journal : for cell and molecular biology **65,** 798-806.

**Johnson, L., Cao, X., and Jacobsen, S.** (2002). Interplay between two epigenetic marks. DNA methylation and histone H3 lysine 9 methylation. Current biology : CB **12,** 1360-1367.

**Josefsson, C., Dilkes, B., and Comai, L.** (2006). Parent-dependent loss of gene silencing during interspecies hybridization. Current biology : CB **16,** 1322-1328.

**Jullien, P.E., Katz, A., Oliva, M., Ohad, N., and Berger, F.** (2006). Polycomb group complexes self-regulate imprinting of the Polycomb group gene MEDEA in Arabidopsis. Current biology : CB **16,** 486-492.

**Jullien, P.E., Susaki, D., Yelagandula, R., Higashiyama, T., and Berger, F.** (2012). DNA methylation dynamics during sexual reproduction in Arabidopsis thaliana. Current biology : CB **22,** 1825-1830.

**Jullien, P.E., Mosquna, A., Ingouff, M., Sakata, T., Ohad, N., and Berger, F.** (2008). Retinoblastoma and its binding partner MSI1 control imprinting in Arabidopsis. PLoS biology **6,** e194.

**Ketting, R.F.** (2011). The many faces of RNAi. Developmental cell **20,** 148-161.

**Kinoshita, T.** (2007). Reproductive barrier and genomic imprinting in the endosperm of flowering plants. Genes & genetic systems **82,** 177-186.

**Kinoshita, T., Miura, A., Choi, Y., Kinoshita, Y., Cao, X., Jacobsen, S.E., Fischer, R.L., and Kakutani, T.** (2004). One-way control of FWA imprinting in Arabidopsis endosperm by DNA methylation. Science **303,** 521-523.

**Kloc, A., Zaratiegui, M., Nora, E., and Martienssen, R.** (2008). RNA interference guides histone modification during the S phase of chromosomal replication. Current biology : CB **18,** 490-495.

**Kohler, C., Wolff, P., and Spillane, C.** (2012). Epigenetic Mechanisms Underlying Genomic Imprinting in Plants. Annual Review of Plant Biology, Vol 63 **63,** 331-352.

**Law, J.A., and Jacobsen, S.E.** (2010). Establishing, maintaining and modifying DNA methylation patterns in plants and animals. Nature reviews. Genetics **11,** 204-220.

**Leblanc, O., Pointe, C., and Hernandez, M.** (2002). Cell cycle progression during endosperm development in Zea mays depends on parental dosage effects. The Plant journal : for cell and molecular biology **32,** 1057-1066.

**Lee, T.J., Pascuzzi, P.E., Settlage, S.B., Shultz, R.W., Tanurdzic, M., Rabinowicz, P.D., Menges, M., Zheng, P., Main, D., Murray, J.A., Sosinski, B., Allen, G.C., Martienssen, R.A., Hanley-Bowdoin, L., Vaughn, M.W., and Thompson, W.F.** (2010). Arabidopsis thaliana chromosome 4 replicates in two phases that correlate with chromatin state. PLoS genetics **6,** e1000982.

**Li, N., and Dickinson, H.G.** (2010). Balance between maternal and paternal alleles sets the timing of resource accumulation in the maize endosperm. Proceedings. Biological sciences / The Royal Society **277,** 3-10.

**Lippman, Z., Gendrel, A.V., Black, M., Vaughn, M.W., Dedhia, N., McCombie, W.R., Lavine, K., Mittal, V., May, B., Kasschau, K.D., Carrington, J.C., Doerge, R.W., Colot, V., and Martienssen, R.** (2004). Role of transposable elements in heterochromatin and epigenetic control. Nature **430,** 471-476.

**Liu, S., Yeh, C.T., Ji, T., Ying, K., Wu, H., Tang, H.M., Fu, Y., Nettleton, D., and Schnable, P.S.** (2009). Mu transposon insertion sites and meiotic recombination events co-localize with epigenetic marks for open chromatin across the maize genome. PLoS genetics **5,** e1000733.

**Lora, J., Hormaza, J.I., and Herrero, M.** (2010). The progamic phase of an early-divergent angiosperm, Annona cherimola (Annonaceae). Annals of botany **105,** 221-231.

**Lu, J., Zhang, C., Baulcombe, D.C., and Chen, Z.J.** (2012). Maternal siRNAs as regulators of parental genome imbalance and gene expression in endosperm of Arabidopsis seeds. Proceedings of the National Academy of Sciences of the United States of America **109,** 5529-5534.

**Madrid, E.N., and Friedman, W.E.** (2010). Female Gametophyte and Early Seed Development in Peperomia (Piperaceae). American journal of botany **97,** 1-14.

**Martienssen, R.A.** (2010). Heterochromatin, small RNA and post-fertilization dysgenesis in allopolyploid and interploid hybrids of Arabidopsis. The New phytologist **186,** 46-53.

**Mol, R., Matthysrochon, E., and Dumas, C.** (1994). The Kinetics of Cytological Events during Double Fertilization in Zea-Mays L. Plant Journal **5,** 197-206.

**Molnar, A., Melnyk, C.W., Bassett, A., Hardcastle, T.J., Dunn, R., and Baulcombe, D.C.** (2010). Small silencing RNAs in plants are mobile and direct epigenetic modification in recipient cells. Science **328,** 872-875.

**Morin, R.D., Aksay, G., Dolgosheina, E., Ebhardt, H.A., Magrini, V., Mardis, E.R., Sahinalp, S.C., and Unrau, P.J.** (2008). Comparative analysis of the small RNA transcriptomes of Pinus contorta and Oryza sativa. Genome research **18,** 571-584.

**Mosher, R.A.** (2010). Maternal control of Pol IV-dependent siRNAs in Arabidopsis endosperm. The New phytologist **186,** 358-364.

**Mosher, R.A., and Melnyk, C.W.** (2010). siRNAs and DNA methylation: seedy epigenetics. Trends in plant science **15,** 204-210.

**Mosher, R.A., Melnyk, C.W., Kelly, K.A., Dunn, R.M., Studholme, D.J., and Baulcombe, D.C.** (2009). Uniparental expression of PolIV-dependent siRNAs in developing endosperm of Arabidopsis. Nature **460,** 283-286.

**Motamedi, M.R., Hong, E.J., Li, X., Gerber, S., Denison, C., Gygi, S., and Moazed, D.** (2008). HP1 proteins form distinct complexes and mediate heterochromatic gene silencing by nonoverlapping mechanisms. Molecular cell **32,** 778-790.

**Muller, K., Bouyer, D., Schnittger, A., and Kermode, A.R.** (2012). Evolutionarily conserved histone methylation dynamics during seed life-cycle transitions. PloS one **7,** e51532.

**Pan, X., Li, Y., and Stein, L.** (2005). Site preferences of insertional mutagenesis agents in Arabidopsis. Plant physiology **137,** 168-175.

**Pontes, O., Costa-Nunes, P., Vithayathil, P., and Pikaard, C.S.** (2009). RNA polymerase V functions in Arabidopsis interphase heterochromatin organization independently of the 24-nt siRNA-directed DNA methylation pathway. Molecular plant **2,** 700-710.

**Rudall, P.J., Eldridge, T., Tratt, J., Ramsay, M.M., Tuckett, R.E., Smith, S.Y., Collinson, M.E., Remizowa, M.V., and Sokoloff, D.D.** (2009). Seed fertilization, development, and germination in Hydatellaceae (Nymphaeales): Implications for endosperm evolution in early angiosperms. American journal of botany **96,** 1581-1593.

**Rusche, L.N., Kirchmaier, A.L., and Rine, J.** (2003). The establishment, inheritance, and function of silenced chromatin in Saccharomyces cerevisiae. Annual review of biochemistry **72,** 481-516.

**Sargant, E.** (1900). Recent work on the results of fertilization in angiosperms. Annals of botany **14,** 689-712.

**Slotkin, R.K., Vaughn, M., Borges, F., Tanurdzic, M., Becker, J.D., Feijo, J.A., and Martienssen, R.A.** (2009). Epigenetic reprogramming and small RNA silencing of transposable elements in pollen. Cell **136,** 461-472.

**Strasburger, E.** (1900). Einige Bemerkungen zur Frage nach der "doppelten Befruchtung" bei den angiospermen. Bot. Zeit. **58,** 294–315.

**Tanurdzic, M., Vaughn, M.W., Jiang, H., Lee, T.J., Slotkin, R.K., Sosinski, B., Thompson, W.F., Doerge, R.W., and Martienssen, R.A.** (2008). Epigenomic consequences of immortalized plant cell suspension culture. PLoS biology **6,** 2880-2895.

**Teixeira, F.K., Heredia, F., Sarazin, A., Roudier, F., Boccara, M., Ciaudo, C., Cruaud, C., Poulain, J., Berdasco, M., Fraga, M.F., Voinnet, O., Wincker, P., Esteller, M., and Colot, V.** (2009). A role for RNAi in the selective correction of DNA methylation defects. Science **323,** 1600-1604.

**Tian, H.Q., Yuan, T., and Russell, S.D.** (2005). Relationship between double fertilization and the cell cycle in male and female gametes of tobacco. Sexual plant reproduction **17,** 243-252.

**Wierzbicki, A.T., Haag, J.R., and Pikaard, C.S.** (2008). Noncoding transcription by RNA polymerase Pol IVb/Pol V mediates transcriptional silencing of overlapping and adjacent genes. Cell **135,** 635-648.

**Zemach, A., McDaniel, I.E., Silva, P., and Zilberman, D.** (2010a). Genome-wide evolutionary analysis of eukaryotic DNA methylation. Science **328,** 916-919.

**Zemach, A., Kim, M.Y., Silva, P., Rodrigues, J.A., Dotson, B., Brooks, M.D., and Zilberman, D.** (2010b). Local DNA hypomethylation activates genes in rice endosperm. Proceedings of the National Academy of Sciences of the United States of America **107,** 18729-18734.

**Zemach, A., Kim, M.Y., Hsieh, P.H., Coleman-Derr, D., Eshed-Williams, L., Thao, K., Harmer, S.L., and Zilberman, D.** (2013). The Arabidopsis Nucleosome Remodeler DDM1 Allows DNA Methyltransferases to Access H1-Containing Heterochromatin. Cell **153,** 193-205.